\documentclass[12pt,a4paper]{article}

\usepackage{epsfig}
\usepackage{exscale}
\def\lsim{\raise0.3ex\hbox{$<$\kern-0.75em\raise-1.1ex\hbox{$\sim$}}}
\def\gsim{\raise0.3ex\hbox{$>$\kern-0.75em\raise-1.1ex\hbox{$\sim$}}}
\newcommand{\be}{\begin{equation}}
\newcommand{\ee}{\end{equation}}
\newcommand{\ba}{\begin{eqnarray}}
\newcommand{\ea}{\end{eqnarray}}

\def\spose#1{\hbox to 0pt{#1\hss}}
\def\ltapprox{\mathrel{\spose{\lower 3pt\hbox{$\mathchar"218$}}
 \raise 2.0pt\hbox{$\mathchar"13C$}}}
\def\gtapprox{\mathrel{\spose{\lower 3pt\hbox{$\mathchar"218$}}
 \raise 2.0pt\hbox{$\mathchar"13E$}}}

\def\ad#1{$\,^{\rm #1}$}
\def\NT{N_\tau}
\def\nt{\ifmmode\NT\else$\NT$\fi}
\def\NS{N_\sigma}
\def\ns{\ifmmode\NS\else$\NS$\fi}

\begin{document}
\begin{titlepage}
\thispagestyle{empty}

\begin{center}
\vspace*{0.8cm}
{{\Large \bf Has \boldmath$T_c$ been measured by heavy ion experiments?  \\}}\vspace*{1.0cm}
{\large F.\ Karsch\ad {1,2} and K.\ Redlich\ad {3,4}}\\ \vspace*{0.8cm}
\centerline {{$^{\rm 1}$}{\em Fakult\"at f\"ur Physik,
    Universit\"at Bielefeld, D-33615 Bielefeld, Germany}}
\centerline {{\large $^{\rm 2}$}{\em Physics Department Brookhaven
National Laboratory, Upton, NY 11973}}
\centerline {{\large $^{\rm 3}$}{\em Institute of Theoretical Physics
University of Wroclaw, PL-50204 Wroclaw, Poland}}
\centerline {{\large $^{\rm 4}$}{\em ExtreMe Matter Institute EMMI, GSI, D-64291 Darmstadt, Germany}}
\protect\date \\ \vspace*{0.9cm}
{\bf   Abstract   \\ } \end{center} \indent

We discuss the role
of cumulants of net baryon number fluctuations
in the analysis of critical behavior in QCD and the study of
freeze-out conditions in heavy ion experiments.
Through the comparison of
the current set of measurements of higher order cumulants
of net baryon number fluctuations with lattice QCD calculations
and results from hadron resonance gas model  we can
learn to what extent freeze-out as, determined by such cumulants,
occurs close to the QCD transition temperature and thus can probe
critical behavior at small values of the baryon chemical potential.
Understanding  how the
relation between freeze-out conditions and the QCD crossover
transition is reflected in properties of the experimentally
determined cumulants is an important prerequisite to search for the 
QCD critical point. We point out that even if perfect
continuum extrapolated lattice QCD results would be available,
it would be inappropriate
to use these observables to extract the value of the QCD transition
temperature at vanishing baryon chemical potential from experimental data.
We furthermore provide indications that a recently performed comparison
of lattice QCD results on cumulants with data from heavy ion experiments
suffer from systematic as well as statistical uncertainties in the
lattice QCD calculations. 
This makes such comparison of lattice QCD calculations with experimental
data at present not useful.

\vfill \begin{flushleft}
\end{flushleft}
\end{titlepage}

\section{Introduction}

The outstanding goal of the on-going RHIC low energy run is to
search for the elusive critical point in the QCD phase diagram.
Promising observables used in this search are higher order
cumulants of net baryon number fluctuations, which have been
advocated to be sensitive to critical behavior in the vicinity
of the chiral phase transition of QCD at vanishing baryon
chemical potential ($\mu_B$) \cite{Ejiri} as well as in the vicinity
of the QCD critical point at $\mu_B>0$ \cite{Stephanov}.

A first comparison of experimental results
on higher order cumulants of baryon number fluctuations \cite{STAR} with
theoretical calculations performed in the framework of lattice
regularized QCD \cite{GG11} as well as hadron resonance gas \cite{redlich}
calculations led to quite good agreement. Recently the analysis of 
\cite{GG11} has been extended in Ref.~\cite{Science} by treating the QCD 
transition temperature as a free parameter that might be constrained 
through the experimental data.
The striking statement made in this publication
is that through comparison of the lattice QCD calculations with
experimental findings for certain cumulants of net baryon number
fluctuations the QCD transition temperature has been determined.
In the following we will discuss if such an observation can
indeed be substantiated.

\section{Lattice Cut-off effects and the continuum limit of QCD}

Cumulants of net baryon number 
fluctuations play an important role when analyzing properties of QCD in
the vicinity of the chiral phase transition temperature,
$T_c (\mu_B)$. They are defined as derivatives of the 
free energy density or pressure ($P$) of a thermodynamic system at 
temperature ($T$) with respect to the baryon chemical potential ($\mu_B$).
The n-th order cumulant is given by
\begin{equation}
\chi_n^{B} = \frac{\partial^n P(T,\mu_B)/T^4}{\partial (\mu_B/T)^n} \; .
\end{equation}
Higher order cumulants are increasingly sensitive to critical behavior.
They diverge in the chiral limit at $T_c (\mu_B)$ as well as at a
possible critical point at $T_c(\mu_B^c)$ for non-zero quark masses.
 
Properties of cumulants at finite temperature and their dependence
on $\mu_B$ have been studied in lattice QCD calculations using
different lattice discretization schemes.
The lattice calculations \cite{GG11,GG}, on which the analysis of cumulants
presented in
\cite{Science} is based, have been performed within a specific
lattice regularization scheme for 2-flavor QCD. Thus, any influence
of the strange quark  on the thermodynamics has been ignored.
The specific lattice
regularization scheme, the standard staggered fermions used in \cite{GG}, 
is known
to be subject to large lattice discretization errors. This is even more
true
on the rather coarse lattice that have been used for the thermodynamics
studies in \cite{Science,GG} and is well known since a long time.
It becomes, for instance, apparent in the high temperature limit of QCD
where discretization errors in the standard staggered fermion scheme on
lattices with temporal extent $N_\tau=4$ and $6$ 
lead to deviations  of bulk thermodynamic observables,
like energy density and pressure, from their known values  in the continuum
limit by more  than 80\% 
 \cite{EoS}. It 
is also known, that calculations with this action performed on such coarse lattices
lead to severe and different  discretization errors in various hadronic 
observables. 
Such errors can to some extent be reduced by forming appropriate ratios 
as it has, for example, been done in the analysis performed in \cite{GG11}
and also previously when lattice results on ratios of cumulants
have been compared to HRG calculations \cite{Ejiri}.
However, discretization errors can not be neglected when one wants to 
determine absolute scales in a lattice calculation.
 
As a consequence of this it is impossible to arrive at
a unique determination of the scale, i.e., the inverse lattice spacing
(cut-off), that could be used to convert lattice 
results to physical units on such coarse lattice as they have been used 
in Ref.~\cite{GG}. Thus, a direct comparison of such
calculations with experiment seems to be excluded right from the
beginning.

\par
The problem with cut-off effects on coarse lattice has  been already  
addressed in Ref.~\cite{GG} which provided
the numerical lattice QCD calculations on which the analysis in
\cite{Science} is based. There, it was noted,  that the quark mass used 
in the calculations  was still to large to reproduce the correct pion to 
rho-meson, ($m_\pi/m_\rho$) and nucleon to rho-meson, $m_{nucleon}/m_\rho$  
mass  ratios.
In fact, the hadron spectrum calculations relevant for setting the
scale in 2-flavor QCD  at finite temperature on lattices with
four and six time-slices have been done long time ago \cite{Gottlieb}.
It is known from theses calculations, that the nucleon to rho-meson mass
ratio suffers from cut-off effects and attains a value of about 1.7 rather
than its physical value of 1.22.
As a consequence, a determination of a lattice scale from either of these
observables would lead to large differences in a determination of the
QCD transition temperature. Using the rho-meson mass to set the scale
on such coarse lattices
gives $T_c = (160 - 167)$ MeV, while the nucleon mass
gives $T_c= (100-110)$~MeV. On the other hand, using a scale
from gluonic observables like the string tension
lead to a substantially larger transition temperature of about
$T_c\sim 190$~MeV \cite{Karsch_DeTar}.
Another way to state this problem is that the nucleon mass would
turn out to be about $1300$~MeV if the physical value of the rho meson
mass would have been used to set the scale. 
This illustrates the severeness of
cutoff effects in the lattice calculations used in \cite{Science} for
the comparison with experimental data. This also illustrates why major
efforts are still being undertaken in lattice calculations to arrive
at a reliable determination of the QCD transition temperature. 
This can be achieved by using
improved actions which reduce the cut-off distortion and by
performing calculations on lattices with smaller
lattice cut-off, i.e., closer to the continuum limit \cite{Fodor,hotQCD}
where a unique scale can be determined and a reliable comparisons
with continuum physics, e.g. experiments, does then become possible.

\section{Are there free parameters in finite temperature Lattice QCD?}

Let us set aside the problem of controlling cut-off effects in lattice
QCD calculations 
and discuss what actually can be
determine through a comparison of lattice QCD results on fluctuations
of net baryon number with heavy ion experiments. As pointed out already 
in the
previous section, the scale for lattice calculations and as such also the
QCD transition temperature is ideally determined through a comparison
of lattice calculations at zero temperature with known spectral properties
of QCD. Nonetheless,
one may take the point of view that one does not want to rely on
experiments that determined the proton and nucleon mass to set the
scale for lattice QCD calculations, but rather wants to use
a heavy ion experiment that determined the freeze-out temperature.
This, however, relies on the fundamental assumption 
that cumulants of
net baryon number fluctuations indeed measure the same freeze-out
temperature that
has been extracted experimentally from ratios of particle abundances 
through a comparison 
with the hadron resonance gas model (HRG) \cite{HRG}.
These assumptions, however, still needed to be justified.
 
Thus, a much more natural approach  is to accept that
lattice QCD calculations (eventually) provide reliable results
for the QCD transition temperature through comparison with
spectral properties of QCD. In fact, the current best estimates
lead to a transition temperature $T_c= (150-160)$~MeV
\cite{Fodor,hotQCD}. It then is much more sensible 
to use the comparison of lattice QCD results with the experimental
data on ratios of cumulants to learn about the freeze-out
conditions probed by these observables. In this way, one can verify 
whether freeze-out, as probed by
ratios of cumulants, indeed happens close to the QCD transition
temperature and therefore can be used to search for the QCD
critical point. One can also learn whether ratios
of cumulants are consistent with freeze-out conditions determined
from particle yields.

All relevant scales needed to compare a lattice calculation with experimental
data thus can reliably be fixed at zero temperature and no additional 
scales are then required to be determined at finite temperature.

\section{What does the comparison of lattice QCD calculations with
measured higher order cumulants tell us?}

We argued in the previous section that even when perfect lattice
calculations, free of cut-off errors, become available one should
not use the comparison between lattice QCD calculations of ratios
of cumulants and their experimental measurements for the determination
of the QCD transition temperature. One should rather accept that
the transition temperature is already provided by the (then perfect)
lattice calculation. The comparison between experiment and theoretical
calculation will then allow us to learn more about the freeze-out
conditions probed by ratios of cumulants and, hopefully, will 
establish them as unique probes of QCD critical behavior in heavy ion 
experiments.

If we follow this approach one may ask whether the analysis
performed in \cite{Science} confirms that
the freeze-out temperature in heavy ion experiments as determined
from higher order cumulants is close to $T_c$.
Such a statement in its
own would be a great success as it would confirm 
that cumulants are indeed the right observables that should be used in
an experimental search for the QCD critical point.
Unfortunately this conclusion can not yet be drawn on the basis
of the analysis presented in \cite{Science}.

To substantiate our skepticism one needs to discuss the 
statistical
significance of the lattice calculations used in \cite{Science}.
One also needs to realize that the 'determination of $T_c$'
presented in \cite{Science} is based on the
significance of a $\chi^2$ analysis. The $\chi^2$-values in this analysis
become large when the ratio between freeze-out temperature ($T_f$)
and QCD transition temperature ($T_c$) becomes small. This is
counter-intuitive
as one would expect that experimental results do agree better
with hadron resonance gas calculations when freeze-out happens further
away from $T_c$. It is not obvious why lattice QCD calculations should
generate discrepancies in such an 'uncritical' region.
Why does a
variation of $T_{f}/T_c$, as it is done in \cite{Science}, then lead
to such a dramatic change in the $\chi^2$ of the fits presented in Fig. 3
of that paper?

First of all one needs to realize that
the large $\chi^2$ values reported in \cite{Science} arise in a region
where $T/T_c$ becomes small, i.e. $T/T_c\ \lsim\ 0.92$. When looking into
the data of Ref.~\cite{GG} it is evident that the statistical
significance of higher order cumulants rapidly decreases when $T/T_c$
is decreased. This is most apparent for the $N_\tau=4$ calculations
reported in \cite{GG05}. Already for this
easier case ($N_\tau=4$) even the fourth order cumulant (and also
the sixth order cumulant \cite{GG}) is negative for $T/T_c \simeq 0.92$.
Both quantities should stay positive in that temperature regime.
On lattices of size $N_\tau=6$ it does become even more difficult
to get the statistics under control. Data for the sixth order
cumulant given in \cite{GG} have a 50\% error at $T/T_c \simeq 0.94$.
In view of this one
may doubt whether the eighth order term is of any use for the Pade
analysis performed in \cite{Science}.

We thus may safely assume that we should not rely on any results
from eighth order cumulants in the comparison of lattice QCD
results with the experimental observable under question. In \cite{Science}
a particular combination of variance ($\sigma$), skewness ($S$)
and kurtosis ($\kappa$), that is proportional to the
ratio of third and fourth order cumulants of net baryon number
fluctuations, has been analyzed
\begin{equation}
\frac{\kappa \sigma}{S} = \frac{\chi_B^{(4)}(T,\mu_B)}{\chi_B^{(3)}(T,\mu_B)}
\; .
\end{equation}
Written in terms of a next-to-leading order Taylor series expansion around
$\mu_B=0$ this quantity is given by
\begin{equation}
\frac{\kappa \sigma}{S} = \frac{T}{\mu_B}\frac{1+\frac{1}{2}
\frac{\chi_B^{(6)}(T,0)}{\chi_B^{(4)}(T,0)} \left( \frac{\mu_B}{T}\right)^2
+ {\cal O}(\mu_B^4)}
{1+\frac{1}{6}
\frac{\chi_B^{(6)}(T,0)}{\chi_B^{(4)}(T,0)} \left( \frac{\mu_B}{T}\right)^2
+ {\cal O}(\mu_B^4)} \; .
\label{ratio}
\end{equation}
Assuming that $\chi_B^{(6)}(T,0)/\chi_B^{(4)}(T,0) > 0$, which will
be the case below $T_c$, one
concludes from this next-to-leading order result that within this
approximation $1< \mu_B\kappa\sigma / ST< 3$.
Here the lower limit corresponds to the HRG result,
$\kappa \sigma / S =1/\sinh (\mu_B/T)\simeq T/\mu_B$, and the upper limit
is reached when the expansion is dominated completely by the next-to-leading
order correction. At this point, of course, one should not trust the expansion
anymore. In fact, the data shown in Fig.~3 of Ref.~\cite{Science} cover
this range of values. For the two data points which generate the large
$\chi^2$ in Fig.~3 of Ref.~\cite{Science}, i.e., the data labeled
with $T_c=180$~MeV and $190$~MeV for $\sqrt{s}=62.4$~GeV the next to leading
order correction
is 100\% and 200\% of the leading order result. Maybe a Pade resummation
is smarter than a direct analysis of the Taylor series. However, in the
absence of any systematic analysis of Pades of different order one may doubt
that this is the case.
\begin{figure}[t]
\begin{center}
\includegraphics[width=7cm]{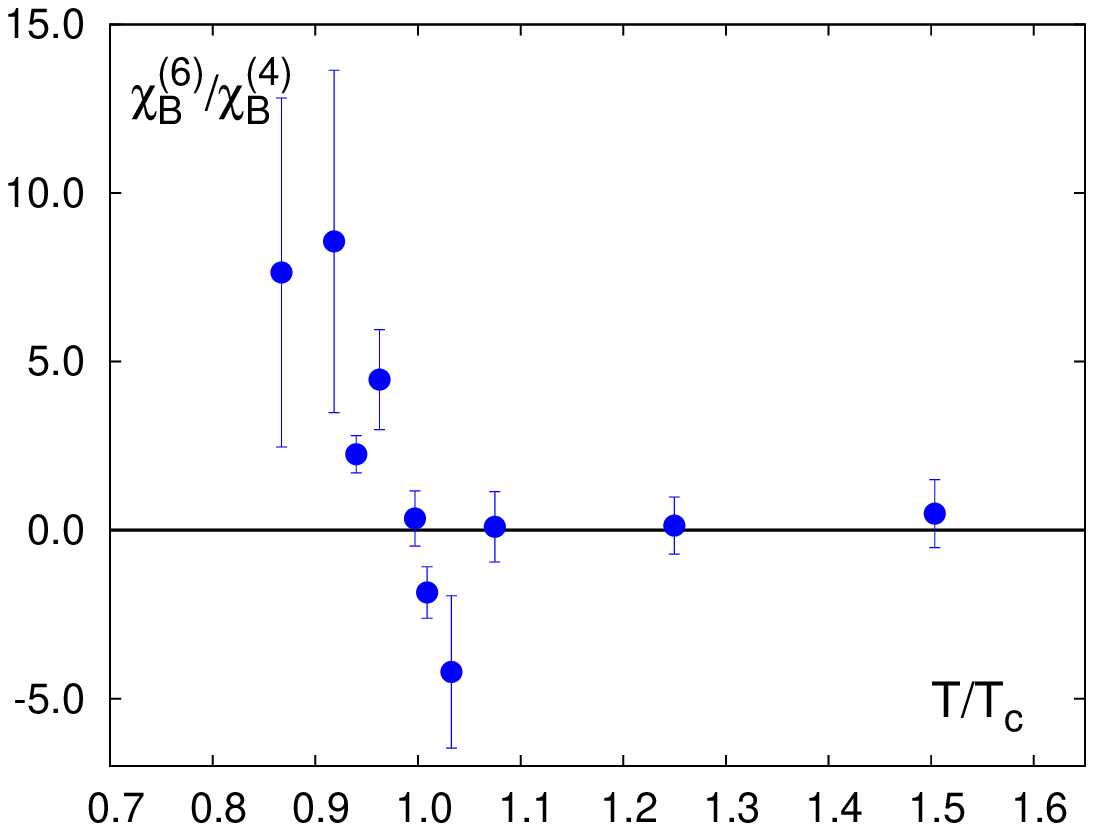}
\includegraphics[width=7cm]{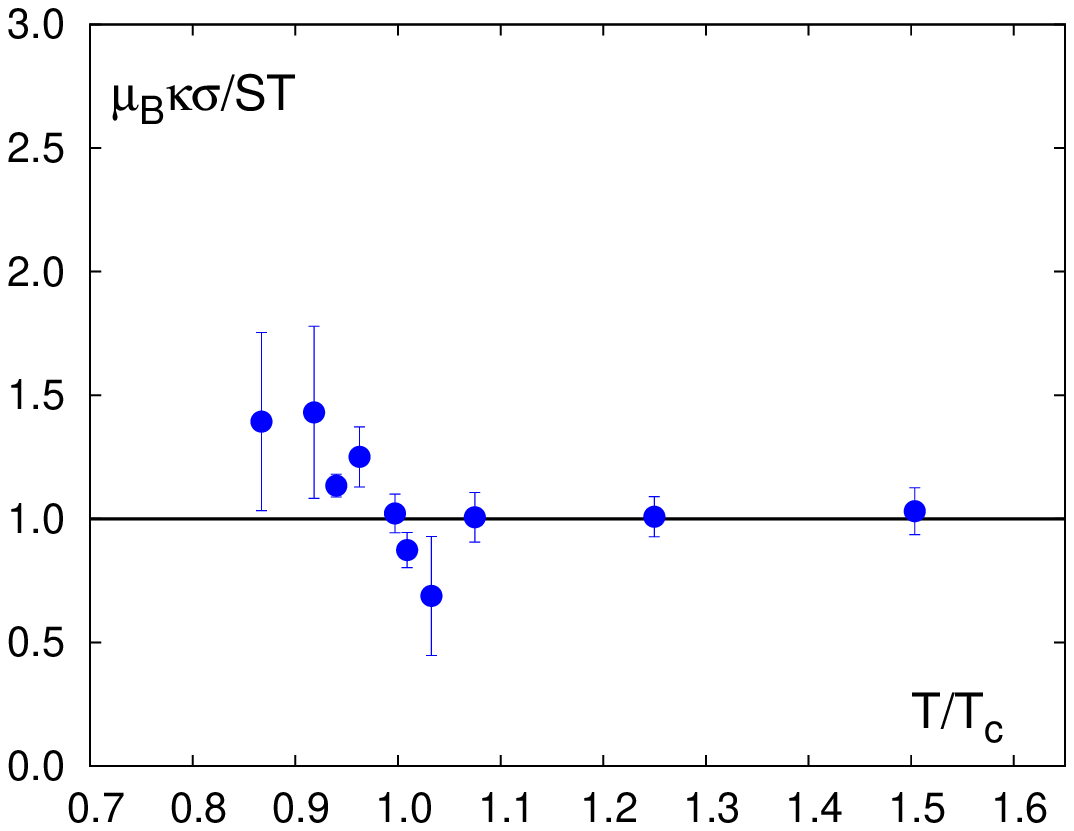}
\end{center}
\caption{\label{fig:p4} The ratio of sixth and fourth order cumulants,
$\chi_B^{(6)}(T,0)/\chi_B^{(4)}(T,0)$, calculated at vanishing baryon
chemical potential using an improved staggered fermion action (p4)
\cite{p4,Schmidt} (left). 
This ratio enters in next-to-leading order in the calculation of
$\chi_B^{(4)}(T,\mu_B)/\chi_B^{(3)}(T,\mu_B)$
and $\mu_B\kappa\sigma / ST$ at the freeze-out temperature.
The right hand figure shows this quantity for
the value of the chemical potential corresponding to the
RHIC low energy run at $\sqrt{s}=62.4$~GeV.
}
\end{figure}

Cumulants up to $6^{th}$ order have also been calculated using an improved
staggered action (p4) on lattices of size $16^3\times 4$ \cite{p4}.
A comparison
of ratios of cumulants with the results from STAR \cite{STAR} have been shown
in Ref.~\cite{Schmidt}.  We have used these data to perform the analysis
done in Ref.~\cite{Science} with numerical results based on calculations
with the p4-action. We use the notation of Ref. \cite{Science},
i.e., we talk about a shift of $T_c$, although we do not like it  
and would rather think of the analysis as probing a shift of the
freeze-out temperature $T_{f}$
relative to a fixed value of $T_c$.

\begin{figure}[t]
\begin{center}
\includegraphics[width=9cm]{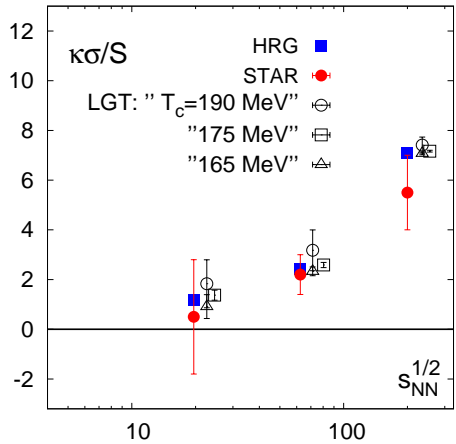}
\end{center}
\caption{\label{fig:ratio} The ratio of fourth and third order cumulants
measured by STAR in the RHIC low energy runs \cite{STAR} and compared to 
lattice results for which the value of $T_c$ has been shifted (see text).
Also shown are results from a HRG calculation \cite{redlich}.
The lattice QCD data have been displaced slightly for better visibility.}
\end{figure}

The crucial data set for the observation of a large variation
in $\chi^2$ in \cite{Science} corresponds to the RHIC low energy run at
$\sqrt{s}=62.4$~GeV.
The value of the baryon chemical potential corresponding to
$\sqrt{s}=62.4$~GeV is $\mu_B\simeq 72.5$~MeV.
Varying $T_f/T_c$ and fixing $\mu_B/T_f$ we calculated
$\mu_B\kappa\sigma / ST$  from Eq.~\ref{ratio} using $T\equiv T_f$. We
never find values for $\mu_B\kappa\sigma / ST$
that become larger than 1.5 (see Fig.~\ref{fig:p4}(right)), although
it is apparent from Fig.~\ref{fig:p4}(left) that the statistical error
on the relevant input variable $\chi_B^{(6)}(T,0)/\chi_B^{(4)}(T,0)$
rapidly increase when $T/T_c\ \lsim\ 0.92$. As $\mu_B\kappa\sigma / ST$
never gets larger than 1.5 the variation of $\mu_B\kappa\sigma / ST$
with $T_f/T_c$ also cannot induce large variations in a $\chi^2$ analysis
as it is shown in the central Fig.~3 of \cite{Science}. We performed
the analogous analysis with the p4 data. The result is shown in
Fig.~\ref{fig:ratio}. From this figure it is evident that there is no
'best choice' for $T_c$. The $\chi^2/dof$ for the difference between
experimental data and the lattice QCD results at the three $T_f/T_c$
values shown in this figure vary between 0.6 and 1.3.

\vspace*{0.2cm}
\section{Conclusions}
We noted that the analysis performed in \cite{Science}, which aimed
at 'Setting the scale for the QCD phase diagram' can not reach this
ambitious goal because the lattice results used in these
calculations suffer themselves from severe cut-off effects and do not
allow for a unique determination of a scale. We have shown, that 
a comparison of current experimental results with another
lattice discretization scheme, which at present suffers from similar
discretization errors, does lead to a different conclusion. This
hints at problems with the statistical significance of the analysis
performed in \cite{Science} as well as with the conclusion that the critical
temperature has been determined by comparing lattice QCD calculations
with results from heavy ion experiments.

Putting aside the issue of actual quality of lattice results used in  
\cite{Science},
we have argued  that the procedure proposed in \cite{Science} to determine 
the critical temperature by comparing lattice QCD results with heavy ion 
data on different cumulants of net baryon number fluctuations 
should not have been done in the first place.
The comparison of lattice QCD results with known spectral 
properties of QCD leads to far more accurate determinations of 
the scale needed to confront lattice QCD calculations with 
results from heavy ion experiments.

\section*{Acknowledgements}
We thank Peter Braun-Munzinger, Bengt Friman, Rajiv Gavai, Edwin Laermann, 
Larry McLerran, Vladimir Skokov, Swagato Mukherjee 
and Nu Xu for discussions and very helpful comments. The work of F.K. was 
supported in part by contract DE-AC02-98CH10886 with the U.S. Department of 
Energy. K.R. acknowledges partial support by the Polish Ministry of Science 
(MEN).

\clearpage
\end{document}